\documentclass[aps,twocolumn,prc,superscriptaddress,showpacs,nofootinbib,floatfix,amssymb,amsfonts,amsmath]{revtex4-1}

\usepackage{graphicx}
\usepackage{epsf}
\usepackage{amsmath,amssymb}
\usepackage{bm}
\usepackage{color}
\usepackage{ulem}
\usepackage{graphicx}
\usepackage{epsf}
\usepackage{amsmath,amssymb}

\newcommand{\bea}{\mbox{$^{10}{\rm Be}$}}
\newcommand{\beb}{\mbox{$^{11}{\rm Be}$}}
\newcommand{\ben}{\mbox{$^{10}{\rm Be}+n$}}
\newcommand{\beq}{\begin{eqnarray}}
\newcommand{\eeq}{\end{eqnarray}}

\begin{document}

\title{ Microscopic cluster study of the $\bea$ and $\beb$ nuclei}

\author{Pierre Descouvemont}
\affiliation{Physique Nucl\'eaire Th\'eorique et Physique Math\'ematique, C.P. 229,
	Universit\'e Libre de Bruxelles (ULB), B 1050 Brussels, Belgium \email{pdesc@ulb.ac.be}}

\affiliation{Yukawa Institute for Theoretical Physics, Kyoto University, 606-8502 Kyoto, Japan}
\author{Naoyuki Itagaki$^2$}

\date{\today}

\begin{abstract}
We use a microscopic multicluster model to investigate the structure of $\bea$ and of $\beb$. These nuclei are
described by $\alpha +\alpha+n+n$ and $\alpha +\alpha+n+n+n$ configurations, respectively, within the
Generator Coordinate Method (GCM). The 4- and 5-body models raise the problem of a large number of generator
coordinates (6 for $\bea$ and 9 for $\beb$), which requires specific treatment. We address this issue by using
the Stochastic Variational Method (SVM), which is based on an optimal choice of the basis functions, generated
randomly. The model provides good energy spectra for low-lying states of both nuclei. We also compute rms radii and 
densities, as well as electromagnetic transition probabilities. We analyze the structure of $\bea$ and of $\beb$
by considering energy curves, where one of the generator coordinates is fixed during the minimization procedure.
\end{abstract}
\maketitle

\section{Introduction}
\label{sec1}
Nuclear clustering is a well-established phenomenon. In particular, the $\alpha$ particle, due to its
large binding energy, is a typical cluster in light nuclei. About 50 years ago, the seminal paper of Ikeda
and his collaborators \cite{ITH68}
showed that $\alpha$-clustering is expected near the $\alpha$ threshold of $4N$ nuclei. This conjecture leads to
the famous Ikeda diagram and was remarkably
confirmed by theory and by experiment. It was even extended to $N\neq Z$ nuclei (see recent reviews in Refs.\ 
\cite{HIK12,DD12,KSK16,KH18}. For example, $\alpha+^{14}$C cluster states are known in $^{18}$O for a long time \cite{DB85}.

Neutron-rich nuclei, in particular in the low-mass region of the nuclear chart, require specific attention. If $\alpha$
clustering is expected, the role of the external neutrons should be addressed by specific methods. A typical example is $^6$He, which requires three-body models to accurately describe the halo neutrons. Beryllium isotopes are particularly
interesting: $^8$Be is the archetype of $\alpha$-cluster nuclei. Although unstable, the ground state is well known to have
a marked $\alpha+\alpha$ cluster structure. Going to heavier Be isotopes require multicluster approaches, where the $\alpha+\alpha$
structure persists, but where the additional neutrons play a role. Multicluster descriptions have been proposed in the past
within the Generator Coordinate Method \cite{De02}, the Antisymmetrized Molecular Dynamics \cite{KH01,Ka02}, the molecular model~\cite{IO00,IOI00,IHO02,It06,IIM08,IIS08} 
or the Resonating Group Method \cite{Ar04,OAS00}. An experimental review of $Z=2-4$ neutron-rich isotopes can
be found in Ref.\ \cite{Fo18}.

In the present work, we aim to investigate the $\bea$ and $\beb$ isotopes within the $\alpha+ \alpha +n+n$ and 
$\alpha +\alpha+n+n+n$ Generator Coordinate Method (GCM). A previous study on $^9$Be \cite{DI18} within this microscopic
approach shows that the $\alpha+\alpha+n$ model is able to reproduce many $^9$Be properties. The main issue
in $\bea$ and $\beb$ is a large number of independent coordinates. In other words, accurate bases require large 
numbers of functions. This problem can be efficiently addressed by the Stochastic Variational Method (SVM), where
a random choice of the basis sets is performed, which permits optimizing the basis \cite{KK77,VS95}. A recent application
of the SVM to $^6$Li, considered as a six-body system, has been performed in Ref.\ \cite{SH19}.

The paper is organized as follows. In sect. \ref{sec2}, we briefly present the microscopic model, and provide some
detail about our use of the SVM. Sections \ref{sec3} and \ref{sec4} are devoted to the $\bea$ and $\beb$ nuclei,
respectively. Concluding remarks and outlook are presented in sect. \ref{sec5}.

\section{The microscopic multicluster model}
\label{sec2}
\subsection{Wave functions}

\begin{figure}[h]
	\centering
	\includegraphics[width=8.5cm]{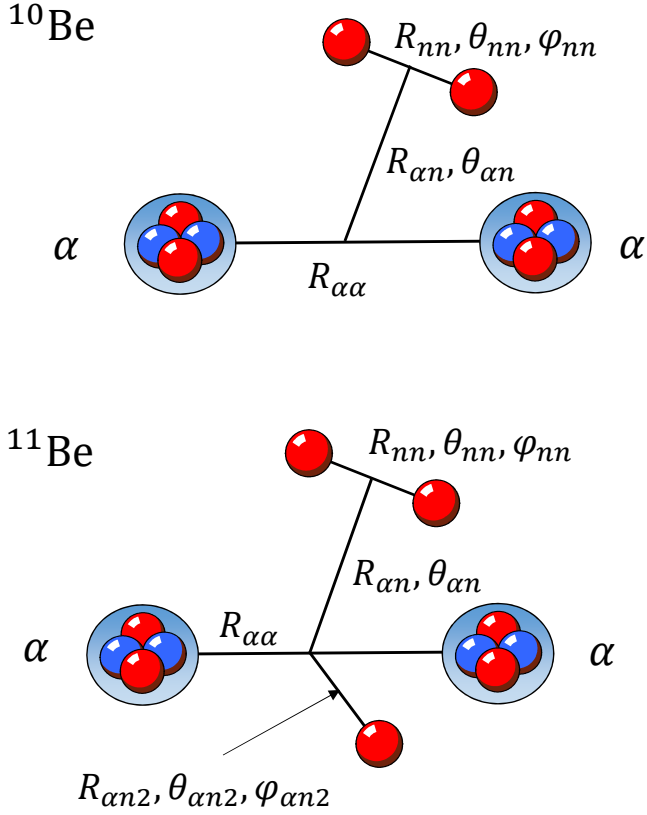} 
	\caption{$\bea$ and $\beb$ configurations with the definitions of the various generator coordinates.}
	\label{fig_config}
\end{figure}

The $\bea$ and $\beb$ isotopes are described in a multicluster model, involving two $\alpha$ particles and two 
or three external neutrons (see Fig.\ \ref{fig_config}).  The Hamiltonian of the system is given by
\beq
H=\sum_{i=1}^A t_i -T_{c.m.} +\sum_{i<j=1}^A v_{ij},
\label{eq1}
\eeq
where $t_i$ is the kinetic energy of nucleon $i$, and $v_{ij}$ a nucleon-nucleon interaction ($A= 10$ or 11).  
The c.m. motion is treated by removing the c.m. kinetic energy $T_{c.m.}$.  We adopt the Minnesota 
interaction \cite{TLT77} as central force, 
complemented by a short-range spin-orbit term \cite{DI18}.  The Minnesota force contains one parameter, 
the admixture parameter $u$, whose standard value is $u=1$, but which can be slightly modified to reproduce 
important properties of the system.  In our work, $u$ is adjusted on the binding energies of $\bea$ or $\beb$.
The Coulomb force is treated exactly.

The wave functions are defined within the GCM \cite{Ho77,Ta81,DD12}. In this microscopic multicluster model, the $^{10,11}$Be isotopes are described by various parameters, referred to as the 
generator coordinates \cite{DD12,Ho77}, which are illustrated in Fig.\ \ref{fig_config}.  As a general statement, 
the number of generator coordinates $ $ increases with the number of clusters $N_C$.  For $^9$Be \cite{DI18}, we have $N_C=3$.  
For $\bea$, this number increases to $N_C=6$.  These parameters are: $R_{\alpha \alpha}$ is the distance between 
the $\alpha$ particles, $R_{\alpha n}$ is the distance between the neutron and $\alpha-\alpha$ center-of-mass, 
$R_{nn}$ is the distance between the external neutrons; three angles ($\theta_{\alpha n},\theta_{nn},\varphi_{nn}$) 
complement the list.  For the $\beb$ nucleus, the additional neutron introduces three new generator 
coordinates ($\theta_{\alpha n2},\theta_{nn2},\varphi_{nn2}$ - see Fig.\ \ref{fig_config}).

For the sake of clarity, we denote as $[R]$ the set of generator coordinates.  In other words, 
$[R]=(R_{\alpha \alpha},R_{\alpha n},\theta_{\alpha n},R_{nn},\theta_{nn},\varphi_{nn})$ for $\bea$ and 
$[R]=(R_{\alpha \alpha},R_{\alpha n},\theta_{\alpha n},R_{nn},\theta_{nn},\varphi_{nn},\theta_{\alpha n2},\theta_{nn2},\varphi_{nn2}) $ 
for $ \beb$.  Of course these large numbers of generator coordinates raise the problem of the basis selection.  This is addressed by the Stochastic Variational Method (SVM) \cite{KK77,VS95} which will be briefly presented in the next subsection.

We first discuss the GCM matrix elements. Let us consider a multicluster wave function as 
\begin{align}
\Phi^{[k]}([R],\Omega)&= {\mathcal A} \phi_{\alpha}(\pmb{S}_1) \phi_{\alpha}(\pmb{S}_2)
\phi_{n}^{k_1}(\pmb{S}_3)\phi_{n}^{k_2}(\pmb{S}_4)
\nonumber \\
{\rm \ (for \ \bea)}, \nonumber \\
&= {\mathcal A} \phi_{\alpha}(\pmb{S}_1) \phi_{\alpha}(\pmb{S}_2)
\phi_{n}^{k_1}(\pmb{S}_3)\phi_{n}^{k_2}(\pmb{S}_4)\phi_{n}^{k_3}(\pmb{S}_5)
\nonumber \\
 {\rm \ (for \ \beb)},
\label{eq2}
\end{align}
where $\pmb{S}_i$ are the locations of the clusters, defined from the generator coordinates $[R]$ and from the 
Euler angles $\Omega$.  In this definition, $\phi_{\alpha}(\pmb{S})$ is an $\alpha$ cluster wave function defined as a 
$(0s)^4$ Slater determinant, and $\phi_{n}^{k}(\pmb{S})$ is a neutron spinor with spin projection $k$.  The $A$-body 
antisymmetrization is taken into account through the operator ${\mathcal A}$.  In Eq.\ (\ref{eq2}), $[k]$ stands 
for the set of spin projections, i.e. $[k]=(k_1,k_2)$ for $\bea$ and $[k]=(k_1,k_2,k_3)$ for $\beb$.  To simplify 
the calculations we assume $k_1=-k_2=1/2$ which represents the dominant component.

In a second step, the basis function (\ref{eq2}) is projected an angular momentum and on parity.  A projected basis function is therefore given by
\beq
\Phi^{JM}_K([R])=\frac{1}{8\pi^2}\int {\mathcal D}^{J\star}_{MK}(\Omega){\mathcal R}(\Omega)
\Phi^{[k]}([R],\Omega) d\Omega,
\label{eq3}
\eeq
where ${\mathcal D}^{J}_{MK}(\Omega)$ is a Wigner function, ${\mathcal R}(\Omega)$ the rotation operator, 
and $K$ is the projection on the intrinsic axis.  The parity projection is performed by superposing another function 
where the center locations are inverted; in a schematic notation, we have
\beq
\Phi^{JM\pi}_K([R])=\frac{1}{2}\bigl(\Phi^{JM}_K([R])+\pi \Phi^{JM}_K(-[R])\bigr).
\label{eq4}
\eeq
Finally, the total wave function of the system is given by superposition of many projected basis functions (\ref{eq4}), 
as
\beq
\Psi^{JM\pi}=\sum_{nK} f^{J\pi}_K([R_n]) \Phi^{JM\pi}_K([R_n]),
\label{eq5}
\eeq
where $f^{J\pi}_K([R_n])$ is the generator function, and is obtained from the diagonalization of the 
Hamiltonian and overlap kernels
\beq
H^{J\pi}_{Kn,K'n'}&=&\langle  \Phi^{JM\pi}_K([R_n]) \vert H \vert  \Phi^{JM\pi}_{k'}([R_{n'}]) \rangle, \nonumber \\
N^{J\pi}_{Kn,K'n'}&=&\langle  \Phi^{JM\pi}_K([R_n]) \vert  \Phi^{JM\pi}_{k'}([R_{n'}]) \rangle.
\label{eq6}
\eeq
These matrix elements, as well as those of  other operators (rms radii, densities, electromagnetic
operators) 
are obtained from three-dimensional integrals over the Euler angles.  As a large number of matrix elements (\ref{eq6})
is necessary, in particular when we optimize the basis with the SVM, a special attention must be paid to 
the efficiency and to the parallelization of the codes.

\subsection{Brief description of the SVM}

The main issue in the present model is to find an optimal set of basis functions, keeping the total number within 
reasonable limits.  The $\bea$ and $\beb$ nuclei are described by 6 and 9 generator coordinates, respectively.  
Obviously, using a standard mesh over each coordinate is not feasible.  This problem can be efficiently addressed 
by using the SVM, widely used in problems dealing with large bases (see, for example, Ref.\ \cite{SH19} for a
recent application).

The SVM has been described in previous references \cite{KK77,VS95}, and we only give here a brief overview.  The SVM 
is based on a random selection of the basis set.  A first set is determined by generating randomly $N_S$ sets and by 
choosing the minimum energy.  The second basis set is obtained in the same way but is coupled to the first set.  
The process is then repeated until the energy remains nearly constant.  Of course, the computer times rapidly
increase when the size of the basis increases.  In practice, we found that $N_S\approx 25-30$ gives a fair convergence.  
The calculations can be tested by repeating the process with another initial set of basis functions.  Obtaining 
close results is a reliable indication that the energy is converged.  This method allows to significantly reduce  
the computer times and memory requirements.

\section{The $\bea$ nucleus}
\label{sec3}

\begin{figure}[h]
	\centering
	\includegraphics[width=8.5cm]{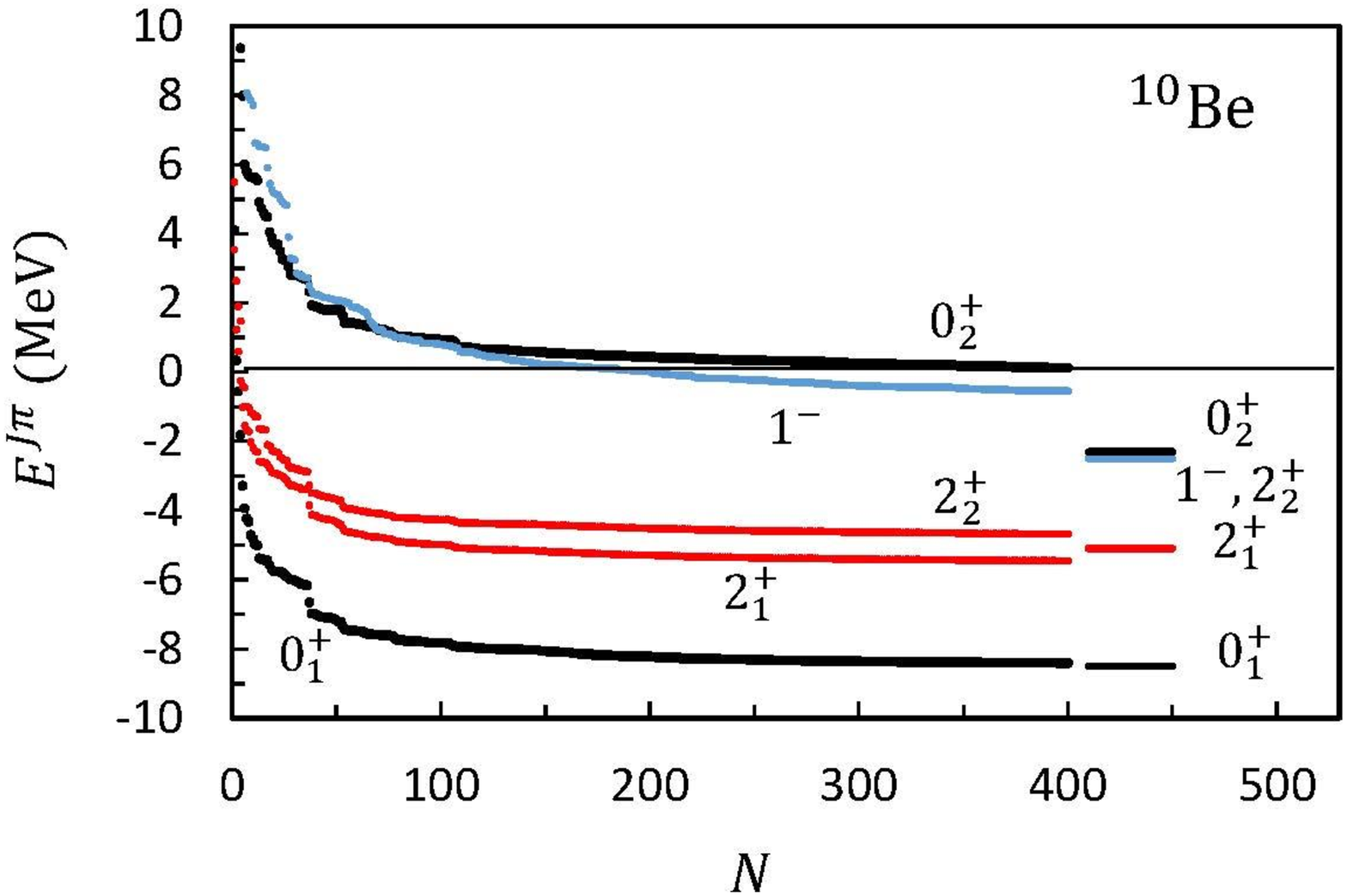} 
	\caption{Convergence of $\bea$ energies with respect to the number of basis functions $N$. The energies are defined
	from the $\alpha+\alpha+n+n$ threshold. Experimental energies of low-lying states are shown on the right of
the figure.}
	\label{fig_be10_conv}
\end{figure}

The energy convergence is illustrated in Fig.\ \ref{fig_be10_conv} for several $J\pi$ values.  We use the Minnesota 
parameter $u=0.973$ which reproduces the experimental binding energy with respect to the $\alpha+\alpha+n+n$ 
threshold ($-8.64$ MeV).  Experimental states are shown on the right part of the figure.  The calculation predicts a $2^+$ excitation energy in excellent agreement with the experiment, although a $2^+_2$ state is found below the experimental
energy.  
The convergence is reasonable with about 400 basis functions.  
We find $0^+_2$ and $1^-$ states whose energies differ by $1-2$ MeV from the experiment.  The $0^+_2$ state is known to 
have an $\alpha+^6$He cluster structure \cite{FBD99}, and its theoretical energy might be slightly improved by 
constraining the random selection to such configurations.  For the negative-parity states, a different $u$ value would permit 
to reproduce more precisely the experimental energy.

\begin{figure}[h]
	\centering
	\includegraphics[width=8.5cm]{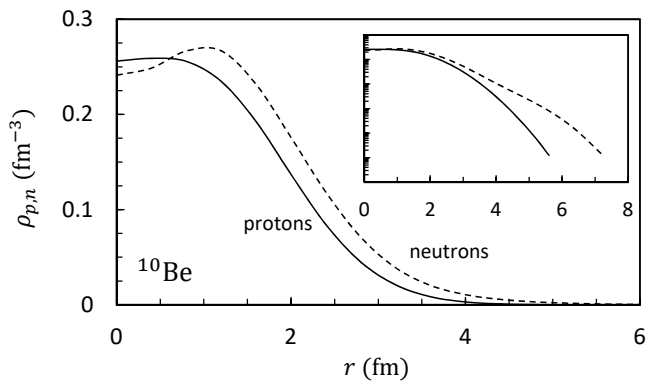} 
	\caption{Proton (solid lines) and neutron (dashed lines) monopole densities of the $\bea$ ground state. The inset shows
	the same densities plotted in a logarithmic scale. }
	\label{fig_dens_be10}
\end{figure}

The ground-state proton and neutron densities of the $\bea$ ground state are presented in Fig.\ \ref{fig_dens_be10}. The 
transition density between an initial state $i$ and a final state $f$ is defined as
\begin{equation}
\rho_{pn}^{i,f} (\pmb r)  =   \langle \Psi^{J_fM_f\pi_f} | \sum_{i=1}^A
\bigl(\frac{1}{2}\pm t_{iz}\bigr)\delta (\pmb r - \pmb r_i) |  \Psi^{J_iM_i\pi_i}  \rangle, 
\label{rho_1}
\end{equation}
where  $\pmb{t}_i$ is the isposin of
nucleon $i$, and the signs "+" and "-" correspond to the neutron and proton densities,
respectively.
These densities are expanded in multipoles \cite{Ka81} as
\begin{equation}
\rho_{pn}^{i,f} (\pmb r)   =\sum_{\lambda } \langle J_f M_f  \lambda M_f-M_i | J_i M_i \rangle 
\rho^{J_j J_i}_{pn,\lambda}(r) 
Y_{\lambda M_f-M_i}^*(\Omega_r),
\label{rho_2}
\end{equation}
and are computed 
as explained in Ref.\ \cite{DI18}.  The monopole ($\lambda=0$) densities of the ground state are normalized such that
\begin{align}
&\sqrt{4\pi} \int \rho_p(r)r^2 dr=Z \nonumber \\
&\sqrt{4\pi} \int \rho_n(r)r^2 dr=N ,
\label{eq7}
\end{align}
where $Z=4$ and $N=6$ for the $\bea$ nucleus.

As expected, the neutron density extends to larger distances, owing to the presence of the external neutrons.  
From these densities, we obtain the rms radii presented in Table \ref{table1}, and compared with the experimental 
data of Ref.\ \cite{THH85}.  The proton radius $\sqrt{<r^2>_p}$ is an excellent agreement with experiment, 
but the matter radius $\sqrt{<r^2>}$ is 
slightly larger.  Notice that the experimental values are partly model dependent.  

With the GCM wave functions, 
we can also compute the E2 transition probability.  Our value is lower than experiment, which suggests that 
an effective charge is necessary.  This is not surprising in neutron-rich nuclei, where polarization effects 
usually require an effective charge which simulates neutron effects.

\begin{table}[!h]
	\caption{$\bea$ properties.}
	\label{table1}
	\centering
	\begin{tabular}{l|ccc}
		\hline
		& GCM & Exp. & Ref. \\
		\hline
 $\sqrt{<r^2>_p}$ (fm) & 2.27 & $2.357\pm 0.018 $ & \cite{NTZ09}\\
 $\sqrt{<r^2>_n}$ (fm) & 2.67 &   & \\
 $\sqrt{<r^2>}$ (fm) & 2.52 & $2.30\pm 0.02$ & \cite{THH85}\\
 $B(E2,0^+\rightarrow 2^+)$ ($e^2$.fm$^4$) & 28.3 & $52\pm 6$ & \cite{TKG04}\\
		\hline
	\end{tabular}
\end{table}

\begin{figure}[h]
	\centering
	\includegraphics[width=8.5cm]{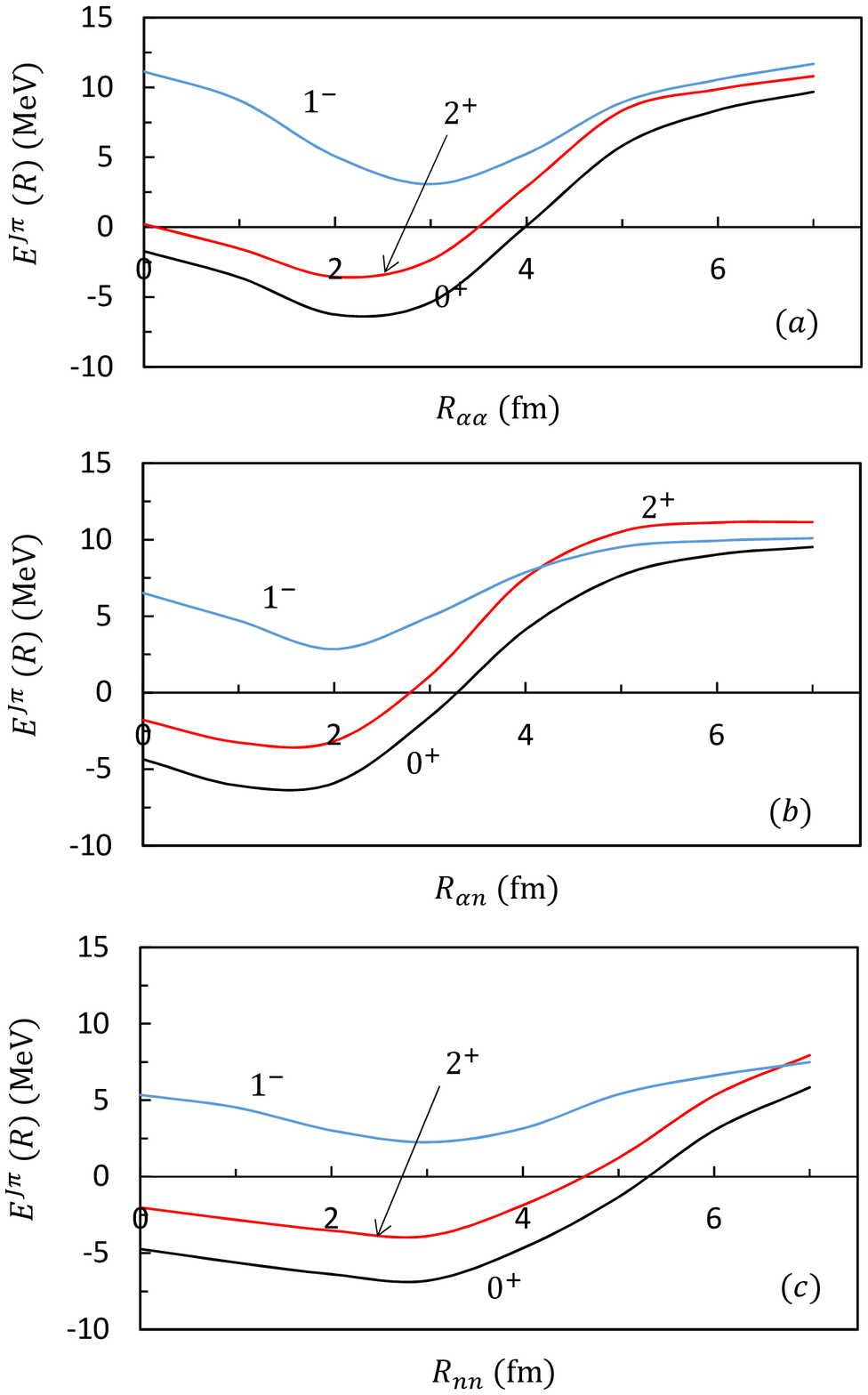} 
	\caption{Energy of the $\bea$ system (with respect to the $\alpha+\alpha+n+n$ threshold) as a function of $R_{\alpha\alpha}$ (a), $R_{\alpha n}$ (b) and $R_{nn}$ (c).}
	\label{fig_be10_r0}
\end{figure}

To have a deeper insight on the $\bea$ structure, we have investigated energy curves, where one of the generator 
coordinate is fixed.  The energy curves are presented in Fig.\ \ref{fig_be10_r0} for $R_{\alpha \alpha},R_{\alpha n}$ and 
$R_{nn}$.  Figure \ref{fig_be10_r0}(a) shows that the minimum of the energy is obtained for 
$R_{\alpha \alpha}\approx 3$ fm, which is lower than in $^9$Be ($\approx 4$ fm) but still significant.  
The $\alpha$-cluster structure is stronger for $J=1^-$, in agreement with the $\alpha+^6$He configuration suggested 
in Ref.\ \cite{FBD99}.  For $R_{\alpha n}$, the minimum is found near $R_{\alpha n}\approx 2$ fm.  
For $R_{nn}$, however, the energy is minimum near $R_{n n}\approx 3$ fm.  This result stresses the importance of 
a 4-body model for $\bea$.  A simple dineutron approximation for the external neutrons would not provide accurate wave functions.

\section{The $\beb$ nucleus}
\label{sec4}
The $\beb$ nucleus has attracted much interest over the last decades, owing to the well-known parity inversion \cite{TU60} and to the 
low binding energy of the ground state. This property makes $\beb$ an ideal example of 
a one-neutron halo nucleus. Many
microscopic studies have been devoted to $\beb$: the GCM \cite{De97,De02}, the AMD \cite{KHD01} and, more recently the No Core Shell Model 
\cite{CNR16} where it is shown an explicit treatment of the $\ben$ cluster structure is necessary to reproduce the
large $B(E1)$ transition probability between the $1/2^+$ ground state and the $1/2^-$ first excited state. In most models,
however, the parity inversion cannot be reproduced with a common interaction. A parity-dependent interaction must be adopted.

In the present work, we aim to investigate the $\beb$ structure in the framework of a multicluster approach.  An
improvement with respect to Ref.\ \cite{De02} is the use of a more efficient method to select the optimal basis, and therefore to get more precise
properties of $\beb$. As mentioned in the introduction, the GCM description of $\beb$ involves 9 generator coordinates, and the
use of the SVM turns out to be quite useful to keep the basis within reasonable sizes.

\begin{figure}[h]
	\centering
	\includegraphics[width=8.5cm]{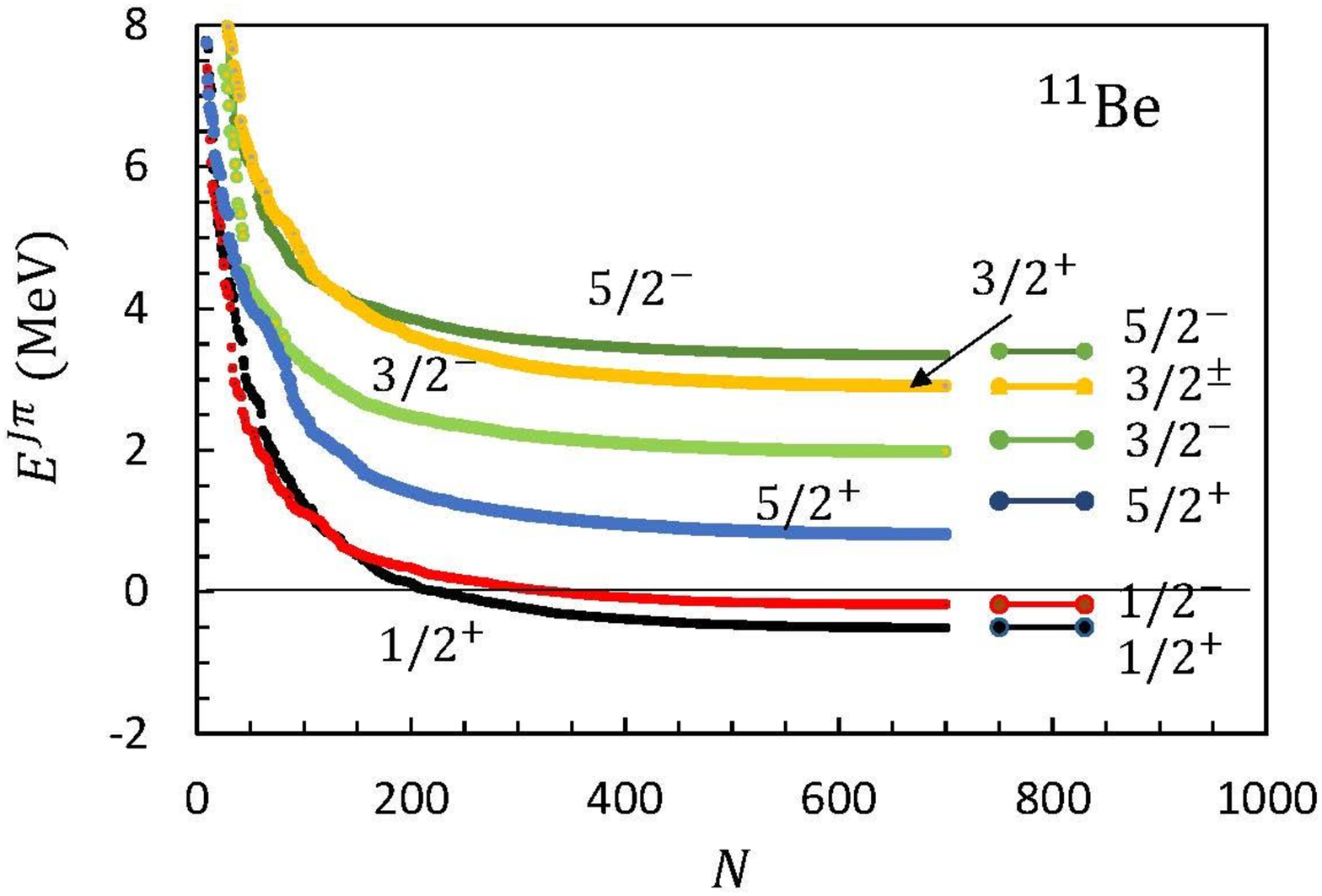} 
	\caption{Convergence of $\beb$ energies with respect to the number of basis functions $N$. The energies are defined
		from the $\alpha+\alpha+n+n+n$ threshold. Experimental energies of low-lying states are shown on the right side of
		the figure.}
	\label{fig_be11_conv}
\end{figure}

We first illustrate the energy convergence of various states in Fig.\ \ref{fig_be11_conv}. The admixture parameter $u$ has been adjusted to
the experimental neutron separation energy. With $u=1.066$ for positive parity and $u=0.893$ for negative parity, we reproduce the energy
of the $1/2^+$ and $1/2^-$ states ($-0.50$ MeV and $-0.18$ MeV, respectively, with respect to the $\ben$ threshold). Reproducing the
experimental binding energies is crucial for the asymptotic part of the wave functions. Figure \ref{fig_be11_conv} shows that a fair
convergence can be achieved with about $600-700$ basis functions. A similar number of basis functions has been employed for the 6-nucleon description of $^6$Li \cite{SH19}. The model not only provides the ground state and the first excited state, but a realistic
description of low-lying resonances is also obtained. For these resonances, the energies are in reasonable agreement with the experiment.

\begin{figure}[h]
	\centering
	\includegraphics[width=8.5cm]{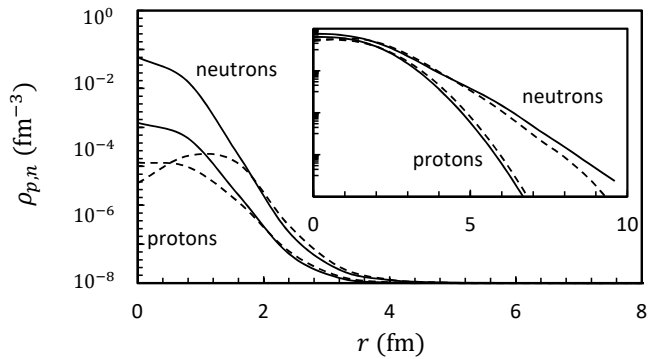} 
	\caption{Proton and neutron  densities of the $\bea$ ground state (solid lines) and of the $1/2^-$ state(dashed lines). The inset shows
	the same densities plotted in a logarithmic scale. }
	\label{fig_be11_dens}
\end{figure}

Figure \ref{fig_be11_dens} presents the $\beb$ proton and neutron densities for the $1/2^+$ and $1/2^-$ states. For
both states, the proton density is rather peaked near the origin. In contrast, the neutron densities extend to large distances. This is a well-known
effect, due to the weak binding energy of the last neutron. The rms radii, obtained from the densities, are displayed in Table
\ref{table2}. The proton and matter radii of the ground state are smaller than the experimental values, a result consistent with
the previous study of Ref.\ \cite{De02}. Most likely, other configurations are necessary to improve the comparison with
experiment. The $B(E1)$ value is also underestimated by the GCM. This was already observed in previous multicluster calculations
\cite{KHD01,De02}, and in the NCSM \cite{CNR16}. The authors of Ref.\ \cite{CNR16} suggest that an explicit account of $\ben$
configurations is necessary to reproduce the large experimental value. 

\begin{table}[!h]
	\caption{$\beb$ properties.}
	\label{table2}
	\centering
	\begin{tabular}{l|ccc}
		\hline
		& GCM & Exp. & Ref. \\
		\hline
		$\sqrt{<r^2>_p}$ (fm) & 1.94 & $2.463 \pm 0.015$ & \cite{NTZ09}\\
		$\sqrt{<r^2>_n}$ (fm) & 2.56   &   & \\
		$\sqrt{<r^2>}$ (fm) & 2.36 & $2.73\pm 0.05$ & \cite{TKY88}\\
		$B(E1,1/2^+\rightarrow 1/2^-)$ (W.u.) & $6.3\times 10^{-3}$  & $0.360\pm 0.033$ & \cite{KKP12}\\
		\hline
	\end{tabular}
\end{table}

\begin{figure}[h]
	\centering
	\includegraphics[width=8.5cm]{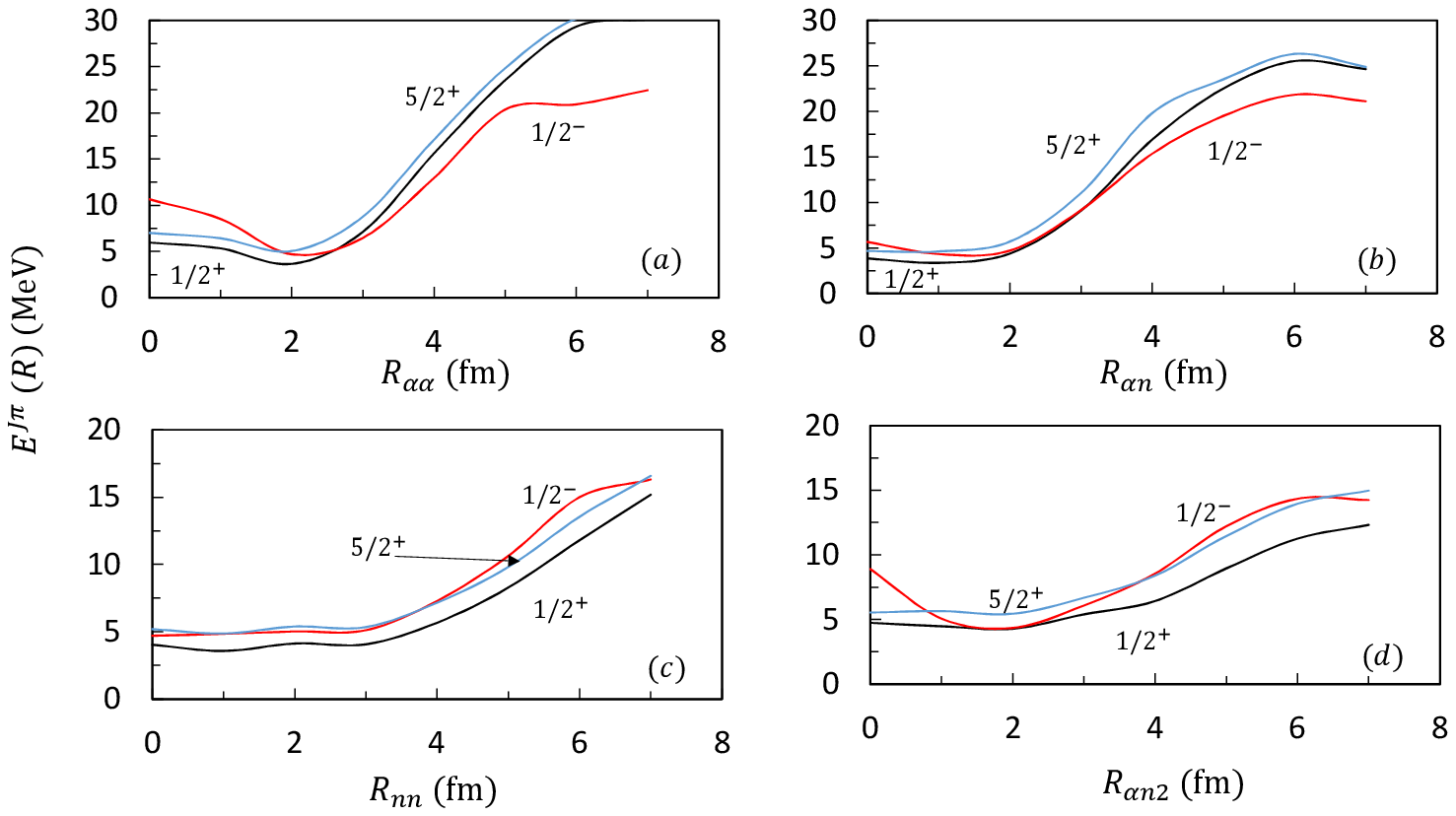} 
	\caption{Energy of the $\beb$ system (with respect to the $\alpha+\alpha+n+n$ threshold) as a function of $R_{\alpha\alpha}$ (a), $R_{\alpha n}$ (b), $R_{\alpha n2}$ (c) and $R_{nn}$ (d).}
	\label{fig_be11_r}
\end{figure}

In Fig.\ \ref{fig_be11_r}, we analyse various energy curves of $\beb$. In each case, a generator coordinate is kept fixed, and the SVM 
is applied to the eight remaining generator coordinates. Although these curves cannot be strictly considered as potentials, they provide
a valuable insight into the structure of $\beb$. The minimum for $R_{\alpha\alpha}$ is found near $R_{\alpha\alpha}\approx 2$ fm, i.e. a value
smaller than in $\bea$. This result confirms that the $\alpha+\alpha$ distance decreases when the number of external nucleon increases \cite{De02}.
For $R_{\alpha n}$, the minimum is rather flat up to $R_{\alpha n}\approx 2$ fm. Large values are therefore unlikely. The $R_{nn}$ dependence is quite interesting: it shows a weak variation of the total energy. Consequently, it is important to include several configurations covering a
wide interval. This conclusion holds for all considered states. The dependence on $R_{\alpha n2}$, i.e. on the distance between the
$\alpha+\alpha$ c.m. and the third neutron, presents a flat minimum around $R_{\alpha n}\approx$ 2 fm.

\section{Conclusion}
\label{sec5}
The main goal of this paper is to investigate the $\bea$ and $\beb$ nuclei within a microscopic multicluster model.
The only adjustable parameter is the admixture parameter $u$, involved in the Minnesota interaction, and fitted on
the binding energy of the ground states. A challenge with many-body approaches is to cope with a large number of generator coordinates or,
in other words, with a large number of degrees of freedom. We have confirmed that the SVM provides an excellent framework
to address this issue. Although computer times are still quite long, they remain within reasonable limits on modern computers.

The multicluster model is based on two $\alpha$ clusters, and 2 or 3 surrounding neutrons. It provides an excellent
description of the low-energy spectrum of both nuclei. In particular, $\beb$ is nicely reproduced, not only for
bound states but also for resonances. The stability of the energies with the number of basis functions 
(see Fig.\ \ref{fig_be11_conv}) shows that we also have a fair description of the continuum.

We have used the GCM wave functions to compute various properties. In particular, the $\bea$ and $\beb$ densities could be
used to determine folding potentials. In $\bea$, the rms radii are in good agreement with the experiment. A small effective
charge should be introduced to improve the agreement for the $B(E2)$ value. The rms radii in $\beb$, however,
are somewhat underestimated, and the $B(E1)$ value is much smaller than the experiment. This is not surprising in
a multicluster model \cite{De02,KHD01}, and even in the NCSM \cite{CNR16}. As suggested in Ref.\ \cite{CNR16}, couplings
to $\ben$ configurations should be introduced explicitly.

Finally, we have analysed the structure of $\bea$ and $\beb$ with the energy curves, where one of the generator coordinates
is fixed. This approach provides a qualitative overview of the nucleus. The $\alpha+\alpha$ clustering decreases
from $\bea$ to $\beb$. It should likely disappear for heavier Be isotopes, such as $^{14}$Be.

\section*{Acknowledgement}
P. D. acknowledges the hospitality of the Yukawa Institute for Theoretical Physics, where most of this work was performed.
This work was supported by the Fonds de la Recherche Scientifique - FNRS under Grant Numbers 4.45.10.08 and J.0049.19.
Computational resources have been provided by the Consortium des Équipements de Calcul Intensif (CÉCI), funded by the Fonds de la Recherche Scientifique de Belgique (F.R.S.-FNRS) under Grant No. 2.5020.11 and by the Walloon Region.
P. D. is Directeur de Recherches of F.R.S.-FNRS, Belgium. 


\end{document}